\documentstyle[12pt]{article}
\newcommand{\gl}{gl(1|1)}
\newcommand{\Y}{Y(gl(1|1))}
\newcommand{\lam}{\lambda}
\newcommand{\Lam}{\Lambda}
\newcommand{\ba}{\begin{eqnarray}}
\newcommand{\na}{\end{eqnarray}}
\newcommand{\ban}{\begin{eqnarray*}}
\newcommand{\nan}{\end{eqnarray*}}
\newtheorem{lemma}{Lemma}

\newtheorem{theorem}{Theorem}
\newtheorem{proposition}{Proposition}
\newcommand{\x}{\otimes}

\begin{document}
\title{\small{\bf REPRESENTATIONS OF SUPER YANGIAN}}
\author{\small R. B. ZHANG\\
\small  Department of Pure Mathematics\\
\small  University of Adelaide\\
\small  Adelaide, S. A.,  Australia}
\date{}
\maketitle

\begin{abstract}
We present in detail the classification of the finite dimensional
irreducible representations of the super Yangian associated with
the Lie superalgebra $gl(1|1)$.
\end{abstract}

\vspace{3cm}

\section{ Introduction}
Many new algebraic structures were discovered in the study
of soluble models in statistical mechanics and quantum field theory.
Amongst them the quantum groups\cite{Drinfeld1}\cite{Jimbo}
and Yangians\cite{Drinfeld2} are particularly interesting.
The former have the structures of quasi triangular
Hopf algebras, admitting universal $R$ matrices which play important
roles in many fields in both mathematical physics and pure
mathematics.  The Yangians have structures closely related to but
distinct from that of the quantum groups. Their representation theory
forms the basis of the quantum inverse scattering method.
Recent research has also revealed that the Yangian structure
is the underlying symmetry of many types of integrable models.

For practical applications, e.g., using the algebraic Bethe Ansatz to
diagonalize Hamiltonians of spin chains, one is primarily
interested in the finite dimensional representations of Yangians.
The systematic study of representations of the Yangians associated
with ordinary Lie algebras was undertaken by Drinfeld\cite{Drinfeld3},
who, using techniques developed in Tarasov's work\cite{Tarasov}
on the $gl(2)$ Yangian, obtained
the necessary and sufficient conditions for irreps to be finite dimensional.
The structures of the  finite dimensional irreps of the $gl(m)$ Yangian
were extensively studied\cite{Reshetikhin1}\cite{Cherednik}\cite{Molev};
representations of $so(n)$ and $sp(2n)$ Yangians and
the twisted Yangians were  studied in \cite{Reshetikhin2} and
\cite{Olshanskii}; and the fundamental irreps of all the Yangians
were investigated by Chari and Pressley\cite{Chari}.

It is also possible to introduce Yangians\cite{Nazarov} and their quantum
analogues\cite{Ymn} associated with the simple Lie superalgebras,
which we will call super Yangians in this paper.
Their structures, and their connections with the Lie superalgebras
and the related quantum supergroups in particular,
were studied by Nazarov\cite{Nazarov} and also in \cite{Ymn}.
However, no attempt has yet been made to develop their
representation theory in a systematic fashion.
As a first step towards developing the representation theory of
super Yangians,
we investigate the finite dimensional irreducible
representations of the Yangian $\Y$ associated
with the Lie superalgebra $gl(1|1)$ in detail.
In section $2$, we construct a $BPW$ type of basis for $\Y$.
In section $3$,  we prove that every finite dimensional irrep
of $\Y$ is of highest weight type, and is uniquely characterized
by the highest weight.   We give the necessary and sufficient
condition for an irrep to be finite dimensional. In section $4$,
we construct an explicit basis for each finite dimensional irrep.

It is relatively well known that the representation theory of
Lie superalgebras and the associated quantum supergroups differs
markedly from that of the ordinary Lie algebras and the corresponding
quantum groups. This is also the case for Yangians and super Yangians,
as we will see in the remainder of this paper.

Probably $gl(1|1)$ is the most extensively studied Lie superalgebra,
because of its connection with $BRST$ and supersymmetry.
Its representations are also rather easy to study, as  only
one and two dimensional irreps exist.
However, we are not so lucky with the super Yangian $\Y$.
As we will see, the study of its representations
is a rather complex problem.
The complication arises primarily from the fact that $\Y$ as
an associative algebra is generated by an infinite number
of generators.  In fact, it is a deformation\cite{deform}
of the universal
enveloping algebra of a subalgebra of the Kac - Moody superalgebra
$\widehat{gl}(1|1)^{(1)}$,  which is spanned by the
infinite number of nonnegative modes.

\section{ BPW theorem }
In order to define the super Yangian $\Y$,  we first explain
some properties of the Lie superalgebra $gl(1|1)$.
Let $\{ E^a_b | a, b = 1, 2\}$ be a homogeneous basis for
a ${\bf Z}_2$ graded vector space over  ${\bf C}$ such that
$E^a_b$ is even if $a=b$, and odd otherwise.
The Lie superalgebra $gl(1|1)$
is this ${\bf Z}_2$ graded vector space endowed with the
following graded commutator
\ba
{[} E^a_b, \ E^c_d\} &=& \delta_b^c E^a_d -
(-1)^{(a+b)(c+d)}  E^c_b.
\na
The vector  module of $gl(1|1)$ is a $2$ - dimensional
${\bf Z}_2$-graded vector space $V$, which has a homogeneous basis
$\{ v^1, v^2\}$ with $v^1$ being even while $v^2$ being odd.
The action of $\gl$  on $V$ is defined by $E^{a}_{b} v^c = \delta_b^c v^a$.
We denote the associated vector representation of
$\gl$ by $\pi$. Then in this basis $\pi(E^{a}_{b})=e^{a}_{b}$, where
$e^{a}_{b}\in End(V)$ are the standard matrix units.

Define the permutation operator $P: V\otimes V\rightarrow V\x V$
by $P(v^a\x v^b)=(-1)^{a b + a + b} v^b\x v^a$. Then explicitly, we have
\ban
P&=&\sum_{a,b=1, 2}e^a_b \otimes e^b_a (-1)^{b+1}.
\nan
It is well known that the following $R$ matrix
\ba
R(u)&=& 1 + { {P}\over {u}},\   \ \ \ \ u\in{\bf C},
\na
satisfies the graded Yang - Baxter equation.

The super Yangian $\Y$ is a ${\bf Z}_2$ graded associative
algebra generated $t^a_b[n]$, $0<n\in{\bf Z}_+$, with some quadratic
relations defined in the following way:
Let
\ban
L(u)&=&\sum_{a, b}(-1)^{b+1}t^a_b(u)\otimes e^b_a,\\
t^a_b(u)&=&(-1)^{b+1}\delta^a_b + \sum_{n=1}^{\infty} t^a_b[n] u^{-n}.
\nan
Then the defining relations of $Y(gl(M|N))$ are
\ba
L_1(u) L_2(v) R_{12}(v-u) &=& R_{12}(v-u)  L_2(v) L_1(u) .\label{def}
\na
Here the grading of the algebra requires some explanation.
The element $t^a_b[n]$ is even if $a=b$, and odd
otherwise. $L(u)$ belongs to the ${\bf Z}_2$ graded vector space
$End(V)\otimes \Y$ and is even.  Equation (\ref{def}) lives
in $End(V)\otimes End(V)\otimes \Y$;  and the multiplication
of the factors on both sides are defined with respect to the
grading of this triple tensor product. More explicitly, we have
\ban
{[} t^{a_1}_{b_1}(u), t^{a_2}_{b_2}(v)\} &=&
{ {(-1)^{\eta(a_1, b_1; a_2, b_2)}}\over {u-v} }
\left[ t^{a_2}_{b_1}(u)t^{a_1}_{b_2}(v)
- t^{a_2}_{b_1}(v) t^{a_1}_{b_2}(u)\right],  \\
{\eta}(a_1, b_1; a_2, b_2)&\equiv&1+ b_2+(b_1+b_2)(a_1+b_2)
+ (a_1+b_1)(a_2+b_2) (mod \ 2);
\nan
or in terms of the generators $t^a_b[n]$,
\ba
{[} t^{a_1}_{b_1}[m], t^{a_2}_{b_2}[n]\}
&=&\sum_{r=0}^{Min(m, n)-1}
\left[ t^{a_2}_{b_1}[r]t^{a_1}_{b_2}[m+n-1-r]
-t^{a_2}_{b_1}[m+n-1-r]t^{a_1}_{b_2}[r]\right]\nonumber\\
&\times& (-1)^{\eta(a_1, b_1; a_2, b_2)}.\label{modes}
\na

A feature of $\Y$ which is not shared by the Yangians associated
with ordinary Lie algebras is that $t^{a}_{b}(u)$, $a\ne b$,
at different $u$ values neither commute nor anticommute. Rather,
\ban
(u-v-1) t^1_2(u) t^1_2(v)&=&-(u-v+1) t^1_2(v) t^1_2(u),\\
(u-v+1) t^2_1(u) t^2_1(v)&=&-(u-v-1) t^2_1(v) t^2_1(u).
\nan
A curious observation is that when the spectral parameters
are assumed to be purely imaginary, $t^{a}_{b}(u)t^{a}_{b}(v)$
and $t^{a}_{b}(v)t^{a}_{b}(u)$($a\ne b$) only differ by a phase.

$\Y$ also admits co - algebraic structures compatible with
the associative multiplication.  We have the
co - unit $\epsilon: \Y\rightarrow{\bf C}$,
$t^a_b[k]\mapsto\delta_{0 k}\delta^a_b(-1)^{a+1}$,
the co - multiplication $\Delta: \Y\rightarrow \Y\otimes \Y$,
$L(u)\mapsto L(u)\otimes L(u)$, and also
the antipode $S: \Y\rightarrow\Y$, $L(u)\mapsto L^{-1}(u)$.
Thus $\Y$ is a ${\bf Z}_2$ graded Hopf algebra.

For the purpose of constructing representations of $\Y$, the
following generalized tensor product structure is more useful:
\ba
\Delta^{(k-1)}_{\alpha}:  \Y&\rightarrow&\Y^{\otimes k}, \nonumber \\
L(u)&\mapsto& L(u+\alpha_1)\otimes L(u+\alpha_2)\otimes ...
\otimes L(u+\alpha_k), \label{comultiplication}
\na
where  $\alpha_1 =0$, and $\alpha_i$, $i=2, 3,  ..., k$,  are
a set of arbitrary complex parameters.  Explicitly, we have
\ban
\Delta^{(k-1)}_{\alpha}(t^a_b(u))&=&\sum_{a_1, ..., a_{k-1}}
(-1)^{1+k+\sum_{i=0}^{k-1} a_i +\sum_{i=1}^{k-1}(a_{i-1}+a_i)(a_i+a_{i+1})}\\
&\times& t^{a_1}_b(u)\otimes t^{a_2}_{a_1}(u+\alpha_2)\otimes ...
\otimes t^a_{a_{k-1}}(u+\alpha_k)
\nan
where $a_0=b$. A further useful fact is the existence of an
automorphism $\phi_f: \Y$ $\rightarrow$ $\Y$ corresponding
to each power series $f(x)=1 + f_1 x^{-1} + f_2 x^{-2}+...$,
which is defined by
\ba
t^a_b(x)&\mapsto&{\tilde t}^a_b(x)=f(x)t^a_b(x).    \label{auto}
\na
As can be easily seen,
the ${\tilde t}^a_b$ satisfy exactly the same relations as
the $t^a_b$ themselves.  For later use, we define
\ban
{\tilde t}^a_b(x) &=&\sum_{k\ge 0}{\tilde t}^a_b[k] x^{-k}.
\nan\\

Let us introduce a filtration on $\Y$.  Define
the degree of a generator $t^a_b[n]$ by $deg(t^a_b[n])=n$, and
require that the degree of a monomial
$t^{a_1 }_{b_1}[n_{1 }]t^{a_2}_{b_2}[n_{2 }]
...t^{a_k}_{b_k}[n_{k}]$ is $\sum_{r=1}^k n_{r}$.
Let $Y_p$ be the vector space over $\bf C$ spanned by monomials of
degree not greater than $p$. Then
\ban
...\supset Y_p\supset Y_{p-1}\supset ... \supset Y_1,\\
Y_p Y_q\subset Y_{p+q}.
\nan

Let $z_1, \ z_2, ..., \ z_k$  be some $t^a_b[n]$'s.
Assume that $Z=z_1  z_2 ... z_k$ has $deg(Z)=p$, then it directly
follows from (\ref{modes}) that
for any permutation $\sigma $ of $(1, 2, ..., k)$,
\ban
z_1  z_2 ... z_k -
\epsilon(\sigma) z_{\sigma(1)} z_{\sigma(2)} ... z_{\sigma(k)}
\nan
belongs to $Y_{p-1}$, where $\epsilon(\sigma)$ is
$-1$ if $\sigma$ permutes
the odd elements in $z_1, \ z_2, ..., \ z_k$ an odd number of times,
and $+1$ otherwise.  In particular, if $t^a_b[n]$ is odd, then
$(t^a_b[n])^2\in Y_{2n-1}$.   Therefore,  given any ordering of the
generators $t^a_b[n]$,  $0<n\in{\bf Z}_+$,   $ a, b=1, 2$,
their ordered products of degrees less or equal to $p$ span $Y_p$,
where the products do not contain factors $(t^a_b[n])^2$ if
$t^a_b[n]$ is odd.

Consider the following ordered products:
\ba
\left(t^2_1[n_1]\right)^{\theta_1}\left(t^2_1[n_2]\right)^{\theta_2}
...\left(t^2_1[n_r]\right)^{\theta_r}\nonumber\\
\times
\left(t^1_1[i_1]\right)^{k_1} \left(t^1_1[i_2]\right)^{k_2}
...\left(t^1_1[i_s]\right)^{k_s}\nonumber\\
\times
\left(t^2_2[j_1]\right)^{l_1}\left(t^2_2[j_2]\right)^{l_2}
...\left(t^2_2[j_t]\right)^{l_t}\nonumber\\
\times
\left(t^1_2[m_1]\right)^{\delta_1}\left(t^1_2[m_2]\right)^{\delta_2}
...\left(t^1_2[m_q]\right)^{\delta_q}, \label{basis}
\na
where $n_1<n_2<  ...< n_r$, similarly for  $i_\alpha$, $j_\alpha$
and $m_\alpha$, and  $k_\alpha,\ l_\alpha\in{\bf Z}_+$,
$\theta_\alpha, \ \delta_\alpha\in\{0, \ 1\}$.
\begin{theorem}:
The elements (\ref{basis}) form a basis of $\Y$.
\end{theorem}
{\em Proof}:
Since the elements of (\ref{basis}) span $\Y$,
we only need to show that they are also linearly independent.
Define $U_p=Y_p/Y_{p-1}$,
then the multiplication of $\Y$ defines a bilinear map
$U_p\otimes U_q\rightarrow U_{p+q}$.
This map extends to $U\otimes U\rightarrow U$,
$U=\oplus_{p=0}^{\infty}U_p$, turning $U$ to an associative algebra.

Introduce the ordinary indeterminates $x_a[n]$,  $a=1, 2$,
$n=1, 2, ...$, and the Grassmanian variables $\zeta^{\pm}[n]$,
$n=1, 2, ...$. Construct the polynomial algebra $G[x, \zeta]$
in these variables, where for the $\zeta$'s, we have
$(\zeta^{\pm}[n])^2=0$.  Now $U$ and $G$ are isomorphic as
associative algebras. This proves that the elements (\ref{basis})
are linearly independent. \\

We denote by $N^+$ the vector space spanned by the elements of
(\ref{basis}) of the form
$\left(t^1_2[m_1]\right)^{\delta_1} ...\left(t^1_2[m_q]\right)^{\delta_q}$,
$\delta_i=0, 1$.
Similarly, we denote by $N^-$ the vector space spanned by all the
$\left(t^2_1[n_1]\right)^{\theta_1} ...\left(t^2_1[n_r]\right)^{\theta_r}$,
and $Y^0$ that spanned by all the
$ \left(t^1_1[i_1]\right)^{k_1} ...\left(t^1_1[i_s]\right)^{k_s}$
$\times \left(t^2_2[j_1]\right)^{l_1}...\left(t^2_2[j_t]\right)^{l_t}$.
It should be pointed out that $N^{\pm}$ and $Y^0$ are not
subalgebras of $\Y$.

\section{Classification of  irreps}
\subsection{Highest weight irreps}
Consider a finite dimensional irreducible $Y(gl(1|1))$ - module $V$.
A nonvanishing vector $v^\Lam_+\in V_0$ is called maximal if
\ba
t^1_2[n] v^\Lam_+=0, & t^a_a[n] v^\Lam_+=\lam_a[n] v^\Lam_+,
&\forall n>0, \ a=1, 2,\label{high}
\na
where $\lam_a[n]\in{\bf C}$.  An irreducible module is called
a highest weight module if it admits a maximal vector.  We define
\ban
\Lam(x)=(\lam_1(x), \lam_2(x)),
& \lam_a(x)=(-1)^{a+1}+\sum_{k>0}\lam_a[k]x^{-k},
\nan
and call $\Lam(x)$ a highest weight of $V$. We wish to show that
\begin{theorem}
Every finite dimensional irreducible $\Y$ - module $V$ contains a
unique(up to scalar multiples) maximal vector $v^\Lam_+$.
\end{theorem}

Define a subspace $V_0=\{v\in V\ | \ t^1_2[m] v=0, \forall n\}$.
The bulk of the proof of the Theorem is contained in
\begin{lemma}
1). $V_0\ne 0$;\\
2). $Y^0$  stabilizes $V_0$; \\
3). For all $v\in V_0$,
\ban
[t^a_a[m], \ t^b_b[n]]v=0, &  a, b=1,2,  \   \   m, n>0.
\nan
\end{lemma}
{\em Proof}: The proof is rather straightforward,
we nevertheless present it here as the results are of crucial
importance for developing the representation theory.

1). Since $t^1_1[1]$ and $t^2_2[1]$ form an abelian Lie algebra,
there exists at least one  nonvanishing $v\in V$ which is
a common eigenvector of $t^1_1[1]$ and $t^2_2[1]$, that is ,
\ban
t^1_1[1]v=\mu_1 v, & t^2_2[1]v=\mu_2 v, &\mu_1, \ \mu_2\in{\bf C}.
\nan
Now any nonvanishing $t^1_2[n_1]t^1_2[n_2]...t^1_2[n_k]v$, $k\ge 0$,
is a common eigenvector of $t^1_1[1]$ and $t^2_2[1]$
with respective eigenvalues $\mu_1 +k$ and $\mu_2 -k$.
Obviously such vectors are linearly independent. Since $V$ is
finite dimensional, $k$ can not increase indefinitely. Thus by
repeatedly applying $t^1_2[m]$'s to $v$ we will arrive at a
$0\ne v_0\in V$ such that
\ban
t^1_2[n]v_0=0, &\forall n\ge 1,\\
t^1_1[1]v_0=\lam_1[1] v_0,
&t^2_2[1]v_0=\lam_2[1] v_0.
\nan
This proves that $V_0$ contains at least one nonzero element.

2). Let $v$ be a vector of $V_0$. We want to prove that
all $t^{a_k}_{a_k}[n_k]t^{a_{k-1}}_{a_{k-1}}[n_{k-1}]...t^{a_1}_{a_1}[n_1] v$,
$a_i=1, 2$, $n_i>0$, $k\ge 0$ are annihilated by $t^1_2[m]$, $m>0$.
The $k=0$ case requires no proof. Assume that all the vectors
$v_{l}=t^{a_{l}}_{a_{l}}[n_{l}]...t^{a_1}_{a_1}[n_1] v$, $l<k$,
 are in $V_0$, then
\ban
&\ & t^1_2[m]t^1_1[n_k]v_{k-1}=[t^1_2[m],  t^1_1[n_k]]v_{k-1}\\
&=&\sum_{r=0}^{min(m, n_k)-1} \left( t^1_1[r] t^1_2[m+n_k-1-r]
 -t^1_1[m+n_k-1-r]t^1_2[r]\right)v_{k-1}=0,\\
&\ & t^1_2[m]t^2_2[n_k] v_{k-1} = -[t^2_2[n_k], t^1_2[m]]v_{k-1}\\
&=& \sum_{r=0}^{min(m, n_k)-1} \left(t^2_2[r] t^1_2[m+n_k-1-r]
                             -t^2_2[m+n_k-1-r]t^1_2[r]\right)v_{k-1}=0.
\nan

3). The following defining relations of $\Y$
\ban
[t^1_1[m], \ t^1_1[n]]&=&[t^2_2[m], \ t^2_2[n]]=0,\\
{[}t^2_2[m], t^1_1[n]]&=&\sum_{r=0}^{min(m, n)-1}
\left(t^2_1[r]t^1_2[m+n-1-r] - t^2_1[m+n-1-r]t^1_2[r]\right),
\nan
and part $2)$ of the Lemma directly lead to part $3)$. \\

\noindent {\em Proof of the Theorem:}
By part $3)$ of the Lemma,   the action of
the $t^a_a[n]$ on $V_0$ coincides with an abelian subalgebra of $gl(V_0)$.
Therefore, Lie's Theorem can be applied, and we conclude that
there exists at least one common eigenvector of all $t^a_a[n]$ in $V_0$.
This proves the existence
of the highest weight vector. Assume $v_+$ and $v_+'$ are two
highest weight vectors of $V$, which are not proportional to each other.
Applying $\Y$ to them generates two nonzero submodules of $V$,
which are not equal. This contradicts the irreducibility of $V$.

\subsection{Construction of highest weight irreps}
To construct irreducible $\Y$ modules, we consider a
one dimensional vector space ${\bf C}v^\Lam_+$.
We define a linear  action of ${\bf Y}^+=Y^0N^+$ on it by
\ba
N^+ v^\Lam_+&=&0,\nonumber\\
t^a_a[n] v^\Lam_+&=&\lam_a[n] v^\Lam_+,\nonumber\\
t^a_a[n]y_0 v^\Lam_+&=&\lam_a[n]y_0 v^\Lam_+, \ \ \ \ \forall y_0\in Y^0.
\label{highest}
\na

 As we pointed out before, $Y^+$ does not form an associate algebra,
thus it does not make sense to ask whether ${\bf C}v^\Lam_+$ is a $Y^+$
module. However, from the proof of the Lemma we can see that
the definition  (\ref{highest}) is consistent with the commutation
relations of $\Y$.
Now we define  the following  vector space
\ban
{\bar V}(\Lam)&=&\Y\otimes_{Y^+}v^\Lam_+.
\nan
Then ${\bar V}(\Lam)$ is a $\Y$ module, which is obviously isomorphic
to $N^-\otimes v^\Lam_+$.

To gain some concrete feel about this module, we explicitly spell out
the action of $\Y$.  Every vector of ${\bar V}(\Lam)$ can be expressed
as $y\otimes v^\Lam_+$ for some  $y\in N^-$. For simplicity,
we write it as $y v^\Lam_+$.
Given any $u\in\Y$, $u y$ can be expressed as a linear sum of the
basis elements (\ref{basis}).  We write
\ban
u y&=& \sum a_{\alpha, \beta} y_-^{(\alpha)} y_0^{(\beta)}
+\sum b_{\mu, \nu, \sigma} y_-^{(\mu)}y_0^{(\nu)}y_+^{(\sigma)},\\
&\ & y_-^{(\alpha)}, y_-^{(\mu)}\in N^-,
\  y_0^{(\beta)}, y_0^{(\nu)} \in Y^0,
\  y_+^{(\sigma)}\in N^+.
\nan
Then
\ban
u (y v^\Lam_+)&=& \sum a_{\alpha, \beta} y_-^{(\alpha)}
y_0^{(\beta)} v^\Lam_+.
\nan

The $\Y$ module ${\bar V}(\Lam)$ is infinite dimensional.
Standard arguments show that it is indecomposable, and
contains a unique maximal proper submodule $M(\Lam)$.  Construct
\ban
V(\Lam)&=& {\bar V}(\Lam)/M(\Lam).
\nan
Then $V(\Lam)$ is an irreducible highest weight $\Y$ module.  \\

Let $V_1(\Lam)$ and $V_2(\Lam)$ be two irreducible $\Y$ modules with
the same highest weight $\Lam(x)$.  Denote by $v^\Lam_{1,+}$ and
$v^\Lam_{2,+}$ their maximal vectors respectively.
Set $W=V_1(\Lam) \oplus V_2(\Lam)$. Then
$v^\Lam_+=(v^\Lam_{1,+}, v^\Lam_{2,+})$
is maximal, and repeated applications of $\Y$ to $v^\Lam_+$ generate
an $\Y$ submodule $V(\Lam)$ of $W$.  Define the module homomorphisms
$P_i: V(\Lam)\rightarrow V_i(\Lam)$  by
\ban
P_1(v_1, v_2)&=&(v_1, 0),\\
P_2(v_1, v_2)&=&(0, v_2), \  \  v_1\in V_1(\Lam), \ v_2\in V_2(\Lam).
\nan
Since
\ban
P_1(v^\Lam_{1,+}, v^\Lam_{2,+})&=&(v^\Lam_{1,+}, 0),\\
P_2(v^\Lam_{1,+}, v^\Lam_{2,+})&=&(0, v^\Lam_{2,+}),
\nan
it follows the irreducibility of $V_1(\Lam)$ and $V_2(\Lam)$ that
$Im P_i = V_i(\Lam)$.  Now $Ker P_1$ is a submodule of $V_2(\Lam)$.
The irreducibility of $V_2(\Lam)$ forces  either $Ker P_1=0$ or
$Ker P_1=V_2(\Lam)$.  But the latter case is not possible, as
$(0, v^\Lam_{2,+})\not\in W$. Similarly we can
show that $Ker P_2=0$.   Hence, $P_i$ are $\Y$ module isomorphisms.

To summarize the preceding discussions, we have
\begin{theorem}
Let $\Lam(x)$ be a pair of Laurent polynomials. Then there exists
a unique irreducible highest weight $\Y$ module $V(\Lam)$ with
highest weight $\Lam(x)$.\\
\end{theorem}

\subsection{Finite dimensionality conditions}
Let  $V(\Lam)$ be an irreducible  highest weight $\Y$ module with  highest
weight $\Lam(x)$. Denote its maximal vector by
$v^\Lam_+$. The vectors $t^2_1[n] v^\Lam_+$, $n=1, 2,...$,
span a vector space, which we denote by $\Omega(\Lam)$.
As an intermediate step towards the classification of the
finite dimensional
irreducible representations of $\Y$, we determine the necessary and
sufficient conditions for  $\Omega(\Lam)$ to be finite dimensional.
The method used here is adopted from Tarasov's work\cite{Tarasov},
but is more algebraic.

First note that if for some $n$, $t^2_1[n+1] v^\Lam_+$
can be expressed as a linear combination of the elements of
$B_n=\{t^2_1[i] v^\Lam_+| i\le n\}$,
then for all $k\ge 0$, $t^2_1[n+1+k] v^\Lam_+$  can be
expressed in terms of the elements of $B_n$.  This fact can be
easily proven using induction on $k$ with the help of the following
obvious relation:
\ban
t^2_1[n+1+k]v^\Lam_+ &=& \lam_1[n+k] t^2_1[1]v^\Lam_+
- \lam_1[1] t^2_1[n+k]v^\Lam_+
-t^1_1[2]t^2_1[n+k]v^\Lam_+.
\nan
Therefore, if $dim\Omega(\Lam)=N$ is finite, then $B_N$ forms a
basis of $\Omega(\Lam)$, and we have
\ba
t^2_1(x)v^\Lam_+=\sum_{i=1}^N \beta_i(x) v_i,
&v_i=t^2_1[i] v^\Lam_+.  \label{lam}
\na
In (\ref{lam}), the $\beta_i$ are power series in $x^{-1}$,
\ba
\beta_i(x) &=&x^{-i} + x^{-n-1}\sum_{k\ge 0}\sigma_i[k] x^{-k},
\na
and the $\sigma_i[k]$ are defined by
\ban
t^2_1[n+1+k]v^\Lam_+ &=& \sum_{i=1}^k]\sigma_i[k] t^2_1[i]v^\Lam_+.
\nan

In order to determine the functional form of the $\beta_i$,
we apply $t^1_1[2]$ to both sides of this equation. By utilizing
the defining relations of $\Y$, we arrive at
the following recursion relation for the $\sigma_i[k]$
\ba
& &\sigma_i[k+1]- \sigma_{i-1}[k]- \sigma_N[k]\sigma_i[0]\nonumber \\
& &=\delta_{i 1} \left\{\lam_1[n+k+1]
-\sum_{j=1}^n\sigma_j[k]\lam_1[j]\right\},\label{master}
\na
where, by definition, $\sigma_0[k]=0$.
Set
\ban
\sigma_i(x)&=& \sum_{k\ge 0}\sigma_i[k] x^{-k},\\
\delta_i(x)&=& 1 + x^{i-N-1}\sigma_i(x).
\nan
Then equation (\ref{master}) can be easily solved, yielding
\ba
\delta_i(x)=\delta_N(x)\left(1 -\sum_{j=i+1}^N\sigma_j[0] x^{j-N-1}\right),
&i=1, 2, ..., N.   \label{de}
\na
Now the $\beta_i$ can be expressed in terms of  the $\delta_j$:
\ba
\beta_i(x)&=&x^{-i}\delta_i(x), \ \ \ i=1, 2, ..., N.\label{be}
\na

This particular form of the $\beta_i(x)$ imposes stringent constraints on the
highest weight $\Lam(x)$. Observe that the $t^a_a[n]$ stabilize the
vector space $\Omega(\Lam)$.  Thus
\ban
t^a_a(x)t^2_1(y)v^\Lam_+&=&\sum_{1\le i, j\le N}(X_a(x))_{i j}\beta_j(y)v_i,
\nan
where $X_a(x)=(-1)^{a+1}+o(x^{-1})$ is an
$N\times N$ matrix in ${\bf C}[[x^{-1}]]$.
Defining relations of $\Y$ lead to
\ban
[t^1_1(x), t^2_1(y)]v^\Lam_+&=&{ {1}\over{x-y}}
\left[\lam_1(y)t^2_1(x)-\lam_1(x)t^2_1(y)\right]v^\Lam_+,
\nan
which in turn yields
\ba
\sum_{1\le j\le N}\left[(X_a(x))_{i j}\beta_j(y) - \lam_1(x) \beta_i(y)\right]
&=& { {1}\over{x-y} } \left[ \lam_1(y) \beta_i(x) -\lam_1(x)\beta_i(y)\right].
\label{bi}
\na
It is always possible to choose  a set of constants  $x_0$ and
$z_i$, $i=1,2,..,N$, such that  $\sum_{i=1}^N z_i\beta_i(x_0)$ does not vanish.
Set $x=x_0$ in (\ref{bi}). Multiply both sides of the resultsnt equation
by $z_i$ then sum over $i$. Some further simple manipulations lead to
the following functional form for $\lam_1$
\ban
\lam_1(x)&=&\sum_{i=1}^N [a_i +b_i x]\beta_i(x),
\nan
where $a_i$ and $b_i$ are complex numbers. Similar calculations yield
\ban
\lam_2(x)&=&\sum_{i=1}^N [c_i +d_i x]\beta_i(x).
\nan
Using equations (\ref{be}) and (\ref{de}), we obtain
\ba
\lam_a(x)&=&\delta_N(x) P_a(x),\nonumber\\
P_a(x)&=& (-1)^{a+1}+ \sum_{k=1}^N p_a[k] x^{-k},  \ \ \ \ p_a[k]\in{\bf C}.
\label{weight}
\na
Therefore, a necessary condition for $\Omega$ to be finite dimensional
is that $\lam_1(x)/\lam_2(x)$ equals the ratio of two polynomials
in $x^{-1}$, which respectively have $\pm 1$ as their constant terms. \\

Let  $\Lam(x)=(\lam_1(x), \lam_2(x))$
satisfy the relation (\ref{weight}), with one of the
$P_a(x)$ of order $N$, and the other of order not greater than $N$.
We further assume that the  $P_a$ do not have common factors.
Construct an irreducible $\Y$ module $V(\Lam)$
with highest weight $\Lam(x)$. Let $v^\Lam_+\in V(\Lam)$
be its maximal vector, and $\Omega(\Lam)$ be  the subspace
of $V(\Lam)$ as defined before.  Using the automorphism (\ref{auto})
with $f(x)=\delta_N^{-1}(x)$,  we have

\ban
\left[{\tilde t}^1_2(x), \  {\tilde t}^2_1(y)\right]v^\Lam_+
&=&{{1}\over{x-y}}\left[P_2(x)P_1(y) - P_1(x) P_2(y)\right]v^\Lam_+.
\nan
It is obvious but rather crucial to observe that $P_2(x)P_1(y) -
P_1(x) P_2(y)$
is divisible by $x^{-1} - y^{-1}$. We have
\ban
{\tilde t}^1_2(x){\tilde t}^2_1(y)v^\Lam_+
&=& y^{-1}\sum_{k=0}^{N-1} q_k(x) y^{-k}v^\Lam_+,
\nan
where $q_k(x)$ are some polynomials in $x^{-1}$.
This equation immediately leads to
\ban
{\tilde t}^1_2(x){\tilde t}^2_1[k]v^\Lam_+ =0,&\forall k>N,
\nan
which  is equivalent to
\ba
{\tilde t}^2_1[k]v^\Lam_+ =0,&\forall k>N.\label{finite}
\na

It follows from equation  (\ref{finite}) and the BPW theorem(
for the generators ${\tilde t^a_b}[k]$ )  that
the module $V(\Lam)$ is spanned by a subset of the following set of
vectors
\ba
{\tilde B}(\Lam)=\left\{v^{\Lam}_+;
\ {\tilde t^2_1}[n_1]{\tilde t^2_1}[n_2] ... {\tilde t^2_1}[n_r] v^{\Lam}_+|
1\le n_1<n_2<  ...< n_r\le N, \ r=1, 2, ..., N\right\}.
\label{B}
\na
Thus the dimension of $V(\Lam)$ is bounded by $2^N$.  To summarize,
we have proved the first part of the following theorem
\begin{theorem}\label{main}
(1).   The irreducible $\Y$  module  $V(\Lam)$ is finite dimensional
if and only if its highest weight $\Lam(x)=(\lam_1(x), \lam_2(x))$
satisfies the following conditions:
\ba
{{\lam_1(x)}\over{\lam_2(x)}}&=&{{P_1(x)}\over{P_2(x)}},\label{condition}
\na
where the $P_a(x)$ are polynomials in $x^{-1}$, which have no common factors,
and $P_a(x)=$ $(-1)^{a+1} +o(x^{-1})$.\\
(2).  Let $N$, called the order of $\Lam(x)$,
be the largest of the orders of the polynomials $P_a(x)$,
then (\ref{B}) forms a basis of $V(\Lam)$.
\end{theorem}

The proof of the second part of this Theorem will be given in
the next section.

\section{Structure  of irreps}
We investigate the structure of the finite dimensional irreps of
$\Y$ in this section. In particular, we will examine the tensor
products of finite dimensional irreps.
We will also  prove the second part of Theorem $\ref{main}$,
thus to obtain an explicit basis for any finite dimensional
irreducible $\Y$ module.

Let $V(\Lam)$ be a finite dimensional irreducible $\Y$ module
with highest weight $\Lam(x)$. Because of the automorphisms $\phi_f$,
we can assume that  $\Lam(x)$ is of the form
\ban
\lam_a(x)&=& P_a(x), \ \ \ \ \ a=1,2,
\nan
where the $P_a(x)$ are polynomials in $x^{-1}$, which do not
 have  common factors. Also,
$P_a(x)$ $=$ $(-1)^{a+1} +o(x^{-1})$.
Let $N$ be the order of $\Lam(x)$.   We have the following
\begin{lemma}
The vanishing of any linear  combination of the vectors
\ban
\left\{v^{\Lam}_+; \ t^2_1[n_1]t^2_1[n_2] ... t^2_1[n_r] v^{\Lam}_+| \
1\le n_1<n_2<  ...< n_r\le N, \ r=1, 2, ..., N\right\}
\nan
would  lead to
\ban
v_-&=& t^2_1[1] t^2_1[2] ... t^2_1[N]  v^{\Lam}_+=0.
\nan
\end{lemma}
{\em Proof}: This follows directly from  the filtration of $\Y$
introduced in section $2$, and the BPW theorem.  To be more
explicit, we consider any vector $v\in V(\Lam)$ of
the form
\ban
v&=&\sum_{1\le n_1<...<n_r\le N} c_{\underline n} t^2_1[n_1] t^2_1[n_2]
... t^2_1[n_r]v^{\Lam}_+.
\nan
It must contain a term $t^2_1[m_1] t^2_1[m_2]  ... t^2_1[m_r]v^{\Lam}_+$
such that $deg(t^2_1[m_1] t^2_1[m_2]  ... t^2_1[m_r])$ is the largest
compared with all other terms of $v$.  We will call this term distinguished.
For our purpose, we can assume that $c_{\{m_1, ..., m_r\} }=1$.
Multiply $v$ by $t^2_1[k_1] t^2_1[k_2] ... t^2_1[k_{N-r}]$,
where $1\le k_1<k_2<...<k_{N-r}\le N$, and $k_i\ne m_s$ for all
$i=1, ..., N-r$, $s=1, ..., r$.
Denote the resultant vector by $\bar v$.  We now apply the BPW theorem to
rewrite $\bar v$ into
\ban
{\bar v}&=& (-1)^\theta v_-\\
&+&\sum_{1\le p_1<...<p_N\le N}
{\bar c}_{\underline p}t^2_1[p_1] t^2_1[p_2] ... t^2_1[p_N] v^{\Lam}_+,
\nan
where the $v_-$ term arises from the distinguished term of $v$,
and $\theta$ may be $0$ or $1$.  It is crucial to
observe that all the other terms in $\bar v$ must have
\ban
deg(t^2_1[p_1] t^2_1[p_2] ... t^2_1[p_N])
&<& deg( t^2_1[1] t^2_1[2] ... t^2_1[N]).
\nan
But this is impossible, unless they all vanish identically.
Hence ${\bar v}=(-1)^\theta v_-$.
Therefore, if $v$ was vanishing, so was $v_-$. \\

Let $V_1(\Lam)$ be an irreducible $\Y$ module of dimension $2^N$
with an order $N$ highest weight $\Lam(x)=(\lam_1(x),  \lam_2(x))$,
where $\lam_a(x)$ are polynomials in $x^{-1}$.
Let  $W(\mu)$ be a two dimensional irreducible $\Y$ module with highest
weight $\mu(x)=(1+\mu_1/x,  -1+\mu_2/x)$.  Introduce a parameter
$\alpha$ such that $1+(\mu_1+\alpha)/x$ does not divide $\lam_2(x)$,
and $-1+(\mu_2-\alpha)/x$ does not divide $\lam_1(x)$. Then
\begin{proposition}
The tensor product $\Y$ module $V(\Lam)\otimes W(\mu)$
is irreducible with respect to the co - multiplication $\Delta_\alpha$
defined by (\ref{comultiplication}).
\end{proposition}
{\em Proof}:
Denote by $v_+$ the maximal  vector of $V_1(\Lam)$ and
set $v_-=t^2_1[1] ... t^2_1[N] v_+$. Let $w_+$ be
the maximal vector of $W$ and define $w_- = t^2_1[1] w_-$.
It is clearly true that
\ban
V(\Lam)\otimes w_-&=& N^{-}(v_+\otimes w_-).
\nan
Consider
\ba
\Delta_\alpha(t^2_1(x))(v_+\otimes w_+)&=&
t^2_1(x)v_+\otimes \left( 1+ {{\mu_1}\over{x+\alpha}}\right) w_+\nonumber\\
&-&\lam_2(x) v_+\otimes {{1}\over{x+\alpha}}w_-. \label{irreducible}
\na
If $\mu_1\ne 0$, we set $x$ to $x_0=-\mu_1 -\alpha$ in the above
equation to arrive at
\ban
v_+\otimes w_- &=&
-\mu_1 (\lam_2(x_0))^{-1} \Delta_\alpha(t^2_1(x))(v_+\otimes w_+)\\
&\in& N^{-1}(v_+\otimes w_+),
\nan
where $\lam_2(x_0)$ is always nonvanishing.
If $\mu_1=0$, terms of orders higher than
$N$ in $x^{-1}$ of equation (\ref{irreducible}) again lead to
\ban
v_+\otimes w_- &\in& N^{-}(v_+\otimes w_+).
\nan
Hence $V_1(\Lam)\otimes w_-\in N^{-}(v_+\otimes w_+)$, and it follows
that $V_1(\Lam)\otimes w_+\in N^{-}(v_+\otimes w_+)$. Therefore
\ba
V_1(\Lam)\otimes W(\mu)&=&N^{-}(v_+\otimes w_+).
\label{tensor}
\na

By considering the equation
\ban
\Delta_{\alpha}(t^1_2(x))(v_-\otimes w_-)&=& t^1_2(x) v_- \otimes
\left( -1 + { {\mu_2 +1}\over{x+\alpha}}\right) w_-\\
&+& \lam_1(x-1) v_-\otimes {{1}\over{x+\alpha}} w_+
\nan
in a similar way we can see that
\ban
v_- \otimes w_+& \in & N^+(v_-\otimes w_-).
\nan
Since $V_1(\Lam)\otimes w_+ = N^+(v_- \otimes w_+)$, we conclude that
\ba
V_1(\Lam)\otimes W(\mu)&=&N^{+}(v_-\otimes w_-).
\label{tensor1}
\na

Equations (\ref{tensor}), (\ref{tensor1}) and  Lemma $2$ together
imply that  $V_1(\Lam)\otimes W(\mu)$ is irreducible  with
respect to the tensor product $\Delta_{\alpha}$. \\

Let $W(\mu^{(i)})$, $i=1, 2, ..., N$, be two dimensional irreducible
$\Y$ modules respectively having highest weights
\ban
\mu^{(i)}(x)&=& \left( 1 +  {{\mu^{(i)}_1}\over {x}},\
  -1 +  {{\mu^{(i)}_2}\over {x}}\right), \\
& & \mu^{(i)}_1 +\mu^{(i)}_2\ne 0.
\nan
Let $\alpha_i$,  $i=1, 2, ..., N$, be a set of complex parameters
such that $\alpha_0=0$, and the polynomials in $x^{-1}$ defined by
\ban
Q_1(x)=\Pi_{i=1}^N\left(1 +  {{\mu^{(i)}_1+\alpha_i}\over {x}}\right), \\
Q_2(x)=-\Pi_{i=1}^N\left(1 - {{\mu^{(i)}_2-\alpha_i}\over {x}}\right),
\nan
do not have common factors.   Then we have
\begin{theorem}\label{major}
The tensor product $\Y$ module
$W(\mu^{(1)})\otimes W(\mu^{(2)})\otimes ... W(\mu^{(N)})$
is irreducible with respect to the co-multiplication
(\ref{comultiplication}), and its highest weight
$\Lam(x)=(\lam_1(x),$ $\lam_2(x) )$
satisfies
\ban
{ {\lam_1(x)}\over{\lam_2(x)} } &=& { {Q_1(x)}\over{Q_2(x)} }.
\nan
\end{theorem}
{\em Proof}:  Define
\ban
\mu(x)&=&\left( 1+ {{\mu_1[1]}\over {x}} +  {{\mu_1[2]}\over {x^2}},  \
-1+ {{\mu_2[1]}\over {x}} +  {{\mu_2[2]}\over {x^2}}\right)\\
&=& \left( \left(1 +  {{\mu^{(1)}_1}\over {x}}\right)
\left(1 +  {{\mu^{(2)}_1+\alpha_2}\over {x^2}}\right),
-\left(1 - {{\mu^{(1)}_2}\over {x}}\right)
\left(1 -  {{\mu^{(2)}_2-\alpha_2}\over {x^2}}\right)\right).
\nan
Obviously $\mu(x)$ is of order $2$.
Construct the irreducible $\Y$ module $U(\mu)$ with highest weight
$\mu(x)$. Let $u_+\in U(\mu)$ be the highest weight vector.
We claim that $u_+$, $t^2_1[1]u_+$, $t^2_1[2]u_+$,  $t^2_1[1] t^2_1[2]u_+$,
form a basis of $U(\mu)$.  To prove our statement, we only need to show that
these vectors are linearly independent.  The independence of
$t^2_1[1]u_+$ and $t^2_1[2]u_+$ follows from the given condition
that $\mu(x)$ is of order $2$.  If $t^2_1[1] t^2_1[2]u_+$ vanished, we would
have
\ban
0&=&t^1_2[1]t^2_1[1] t^2_1[2]u_+ \\
&=&(\mu_1[1]+\mu_2[1])t^2_1[2]u_+ + (\mu_1[2]+\mu_2[2])t^2_1[1]u_+,
\nan
which implies
\ban \mu_1[1]+\mu_2[1]&=&\mu_1[2]+\mu_2[2] =0. \nan
This would force the order of $\mu(x)$ to be $0$, contradicting
the given conditions.  Therefore our claim is indeed correct.

Now we use induction to prove the theorem. Consider the case $N=2$
first.    By comparing the dimensions of
$W(\mu^{(1)})\otimes W(\mu^{(2)})$ with $U(\mu)$,  we can see that
these two modules coincide up to an appropriate $\phi_f$ automorphism
of $\Y$.  Assume that the  Theorem is valid for $N=k-1$, then it immediately
follows the Proposition that the Theorem is correct for $N=k$ as well.\\

\noindent
{\em Proof of the second part of Theorem $\ref{main}$}: It follows from
Theorem $\ref{major}$ as a  corollary.

\vspace{4cm}
\noindent
{\bf Acknowledgements}: This work was completed while I
visited the Department of Physics, National Chung Hsing
University, Taiwan, China.
I wish to thank Professor H. C. Lee and the other colleagues
in the Department for their hospitality.   Financial support
from both the ARC and NSC is gratefully acknowledged.

\pagebreak

\end{document}